\documentclass{article}
\usepackage{spconf,amsmath}
\usepackage{amssymb,latexsym, fancyhdr, epsfig}

\newcommand{\w}{\mathbf{w}}

\newcommand{\x}{\mathbf{x}}

\newcommand{\bx}{\mathbf{\bar x}}
\newcommand{\y}{\mathbf{y}}
\newcommand{\n}{\mathbf{n}}

\newcommand{\bdalpha}{\boldsymbol \alpha}

\title{Audio Watermarking over the Air With Modulated Self-Correlation}
\name{Yuan-Yen Tai, Mohamed F. Mansour}
\address{Amazon Inc., USA}
\begin{document}
\ninept
\maketitle
\begin{abstract}
We propose a novel audio watermarking system that is robust to the distortion due to the indoor acoustic propagation channel between the loudspeaker and the receiving microphone. The system utilizes a set of new algorithms that effectively mitigate the impact of  room reverberation and interfering sound sources without using dereverberation procedures. The decoder has low-latency and it operates asynchronously, which alleviates the need for explicit synchronization with the encoder.  It is also robust to standard audio processing operations in legacy watermarking systems, e.g., compression and volume change. The effectiveness of the system is established with a real-time system under  general room conditions.
\end{abstract}
\section{Introduction}
\label{sec:intro}

In most existing audio watermarking scenarios in the literature, the audio signal stays in the digital domain between the encoder and the decoder. This is a typical situation in  digital right management of audio distribution, where the watermarking decoder is invoked prior to media playback \cite{swanson1998audio, hua2016twenty}. Recently, there has been growing interest in audio watermarking that survives indoor acoustic propagation, e.g., for second-screen applications \cite{cesar2009leveraging}. In this scenario, the watermarked audio is played through a consumer loudspeaker after the encoder, propagates through an indoor acoustic channel, picked by a consumer microphone (usually in another device) before passing to the watermark decoder. This scenario poses a set of new challenges that were not encountered in legacy audio watermarking:
\begin{itemize}
 \item Room reverberation, which introduces time and frequency smearing of the audio content \cite{kuttruff2016room}.
 \item Time/frequency drift between the encoder and decoder due to different system clocks. 
 \end{itemize}

The relevant work in the literature has treated these two challenges rather separately, and frequently at the cost of less robustness to standard audio processing operations.  For example, few audio watermarking systems have been designed to withstand desynchronization between the encoder and decoder \cite{mansour2003time, pun2013robust, wang2006novel, xiang2014patchwork, nadeau2017audio, del2011audio}. This robustness could be achieved  through using features that are robust to local time-scale variations \cite{mansour2003time}, or deploying a special synchronization mechanism (through time-warping like procedure) at the decoder \cite{xiang2014patchwork, nadeau2017audio}. On the other hand, some earlier works have focused on the reverberation impact while assuming perfect synchronization  \cite{del2011audio, doubledct}. In \cite{del2011audio}, a special filter bank with a long symbol interval is used, and the watermark is embedded in the specific time-frequency cells that are robust to expected operations. The synchronization has not been explicitly addressed, rather general guidelines from wireless communication systems were described. In this work, we develop an end-to-end audio watermarking system that addresses these challenges under practical computation and latency constraints.

In particular, we develop a novel audio watermarking system that is robust to both reverberation and desynchronization  as well as standard audio processing operations. The encoder embeds a spread-spectrum watermark in successive short blocks of the host audio, and the watermark at each block is modulated with a binary $\pm 1$ sequence to improve the detection and suppress host signal correlation. The encoder resembles standard audio watermarking systems, therefore, it inherits  their good properties, e.g., imperceptibility of the watermark, and robustness to standard signal processing operation such as audio coding and filtering. The decoder applies a modulated \emph{self-correlation} of successive blocks rather than the standard matched filter that uses cross-correlation with the embedded watermark (which requires perfect synchronization and knowledge of the acoustic channel at the decoder). 
Although self-correlation is not the optimal detector from detection theory perspective, it effectively and blindly mitigates the impact of both reverberation and desynchronization at a low-cost in both computation and latency, which enables real-time embedded implementation. The degradation in the detection performance is shown to be small for practical use cases.

The following notations are used throughout the paper. A bold lower-case letter denotes a column vector. $v_k$ denotes the $k$-th element of $\mathbf{v}$. $\mathbf{x}$ and $\tilde{\mathbf{x}}$  denote respectively, host and watermarked signal at the encoder, while $\mathbf{y}$ denotes the watermarked signal at the decoder. $\langle.,.\rangle$ denotes the inner product.  Additional notations are introduced when needed.

\section{Background}
\subsection{Spread-Spectrum Watermarking}
In the following, we assume that the watermark is embedded in selected DCT coefficients of audio blocks. Spread-Spectrum watermarking procedure has the general form \cite{swanson1998audio, cox1996secure, kirovski2003spread}
\vspace{-0.1cm}
\begin{equation}
    \tilde{\mathbf{x}}  = \mathbf{x} + \eta \mathbf{w}  \label{eq:wm_embed}
\end{equation}
where $\eta$ is the watermark strength (which controls the audibility of the watermark). If $\mathbf{y}$ is the received signal in the DCT domain, then the standard spread-spectrum decoder uses cross correlation of the form
\begin{equation}
\rho = \langle \mathbf{y}, \mathbf{w} \rangle
\end{equation}
In the additive noise case $\mathbf{y} = \mathbf{x} + \eta \mathbf{w} + \mathbf{n}$ (where $\mathbf{n}$ is the noise component), and 
\begin{equation}
\langle \mathbf{y}, \mathbf{w} \rangle = \langle \mathbf{x}, \mathbf{w} \rangle + \langle \mathbf{n}, \mathbf{w} \rangle + \eta \|\mathbf{w}\|^2 \label{eq:cc}
\end{equation}
If the watermark is not correlated with the signal nor the noise, then both $ \langle \mathbf{x}, \mathbf{w} \rangle$ and $ \langle \mathbf{x}, \mathbf{n} \rangle$ vanish and $\rho$ becomes proportional to the watermark energy.  
At the detector,  $\rho$ is compared by a predetermined threshold, $\gamma$. If $\rho \ge \gamma$, then the watermark is detected at the decoder; otherwise, it is not detected.
\vspace{-0.2cm}
\subsection{Acoustic Channel Model}
The acoustic propagation channel has few sources of distortions: clock drift between the encoder and the decoder, sampling rate difference, loudspeaker behavior, room reverberation, and analog-to-digital and digital-to-analog distortion. The microphone impact is usually ignored because of its flat response over the frequencies of interest. 
The clock drift is measured in parts-per-million (ppm), and consumer-grade system clocks can have up to few hundreds ppm error. If the clock drift is $100$ ppm, then at $48$ kHz sampling frequency, the effective sampling frequency is $48000 \pm 4.8$ Hz, which results in a time shift of up to 4.8 samples every second, and also a slight frequency shift. If an explicit synchronization procedure is used, then this clock drift must be estimated and corrected (through PLL-like systems \cite{PLLbook}). The operating sampling rate difference could be mitigated by standardizing the sampling frequency at which the watermark is embedded or detected.
The other distortions can be broadly modeled as a slowly time-varying channel with additive noise similar to fading channels in wireless communication: 
\vspace{-0.1cm}
\begin{equation}
y(t) = \sum_\tau h^{(t)} (\tau)  \tilde{x} (t-\tau) + n(t)
\vspace{-0.2cm}
\end{equation}
where $\{h^{(t)} (\tau)\}$ is the time-varying impulse response, and $n(t)$ is the additive noise. In the frequency-domain, we have
\begin{equation}
{y}^{(t)}(\omega_k) = h^{(t)}(\omega_k)\tilde{x}^{(t)}(\omega_k) + {n}^{(t)}(\omega_k)
\vspace{-0.2cm}
\end{equation}
where ${y}^{(t)}(\omega_k)$ is the frequency response of $y$ at audio frame $t$, and similarly for ${x}^{(t)}(\omega_k) $ and ${n}^{(t)}(\omega_k)$.
If the channel change is slow compared to the watermark length, then in vector-form, each DCT block can be represented as \cite{martucci1994symmetric}
\vspace{-0.2cm}
\begin{equation}
    {\mathbf{y}} = \tilde{\mathbf{x}} \odot {\boldsymbol \alpha} + {\mathbf{n}}  \label{eq:chan-model}
\vspace{-0.1cm}
\end{equation}
where $\odot$ denotes element-wise vector multiplication, and $\boldsymbol \alpha$ is the channel representation in the DCT domain. If $\alpha_k$ changes amplitude and sign with the frequency index $k$ (which is the typical case), then spread-spectrum based audio watermark detection would fail. To see this, consider the cross-correlation factor in this case (assuming perfect synchronization) 
\vspace{-0.2cm}
\begin{equation}
\begin{aligned}
    \langle \mathbf{y}, \mathbf{w} \rangle &\approx  \langle \mathbf{x} \odot {\boldsymbol \alpha}, \mathbf{w} \rangle + \eta \ \langle \mathbf{w} \odot {\boldsymbol \alpha}, \mathbf{w} \rangle \\
    &\approx \eta \sum_k \alpha_k \ | w_k | ^2  \label{eq:cc-reverb}
\end{aligned}
\vspace{-0.2cm}
\end{equation}
If the sign of $\alpha_k$ changes with $k$, then the cross-correlation becomes close to the noise level, and detection fails.
In this case, the optimal detector requires knowledge of the channel at the receiver, which is the common approach in wireless communication. The estimation is performed by transmitting a known pilot signal at the start of each frame, which is used for system identification at the receiver. The channel estimation procedure requires perfect synchronization, and it is an expensive procedure in both computation and latency.

\begin{figure}
\includegraphics[scale=0.15,angle=0]{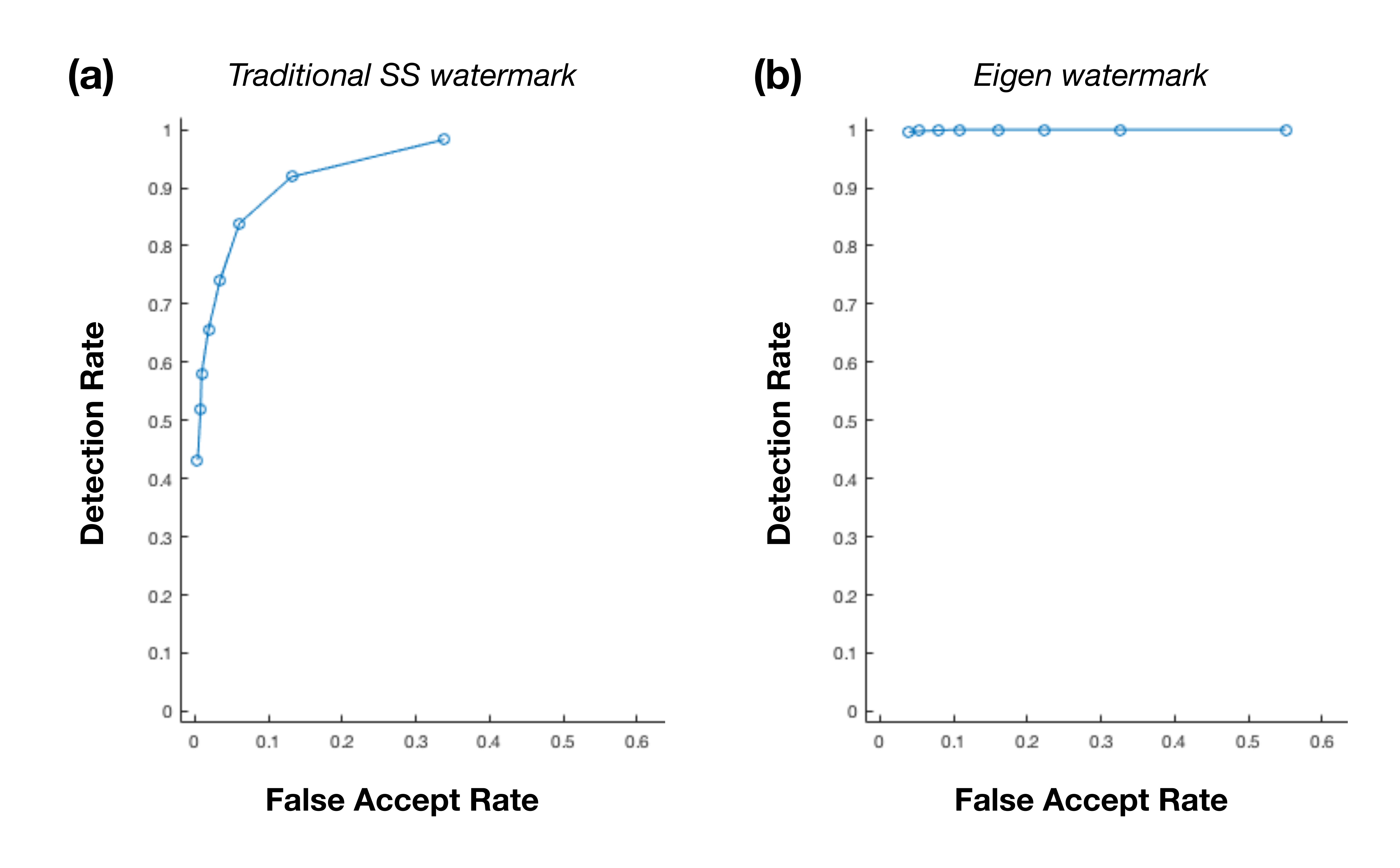}
\vspace{-0.5cm}
\caption {
The ROC metric of `{\it traditional SS watermark}' and `{\it eigen watermark}'.
Each data point is averaged over 1000+ audio pieces that embedded with the watermark with simulated reverberation effect added to the watermarked audio source.
}\label{fig:ROC_eig} 
\vspace{-0.4cm}
\end{figure}
\vspace{-0.1cm}

\section{Watermarking System}
\vspace{-0.1cm}
\subsection{Watermark Design}
If $\mathbf{H}$ is a full-rank symmetric matrix of size $\kappa$, then its eigenvectors $\{\mathbf{v}_i\}$ are real and constitute a set of orthonormal basis for $\mathcal{R}^\kappa$ \cite{matrixbook}. Let the watermark $\mathbf{w}$ be chosen as one of the eigenvectors, e.g., ${\mathbf{v}}_1$ (without loss of generality). The host signal block, $\mathbf{x}$ in \eqref{eq:wm_embed}, can be expressed as
\begin{equation}
\mathbf{x} = \sum_l a_l \ \mathbf{v}_l
\vspace{-0.1cm}
\end{equation}
where $a_l = \langle \mathbf{x}, \mathbf{v}_l \rangle$. In this case, the cross-correlation between the host signal and the watermark becomes
$\langle \mathbf{x}, \mathbf{w} \rangle = a_1$. 
This constitutes the detection noise floor in the noiseless case (i.e., when $\mathbf{n} = 0$ in \eqref{eq:cc}). To completely remove this noise floor in the noiseless case, the host signal is slightly modified to remove the projection component of the host signal onto the watermark subspace. If we choose $\mathbf{w} = \mathbf{v}_1$, then the watermark embedding equation in \eqref{eq:wm_embed} is modified to
\begin{equation}
\begin{aligned}
    \tilde{\mathbf{x}}  =& \;\mathbf{x} - \langle \mathbf{x}, \mathbf{v}_1 \rangle \mathbf{v}_1 + \eta \mathbf{v}_1 \\
                        =& \;\bx + \eta \mathbf{v}_1 \;\;\;(\text{where, } \bx \triangleq \x-\langle \x, \mathbf{v}_1 \rangle \mathbf{v}_1)
    \label{eq:eig_wm_embed}
\end{aligned}
\end{equation}
In Fig.~\ref{fig:ROC_eig},  we show that Eigen Watermarking significantly improves the ROC of the detector and has a good operation point against the simulated reverberation effect.

In order for the embedding algorithm to be robust to standard audio processing operations, e.g., filtering and compression, the embedding is restricted to mid-range DCT coefficients, i.e., in the range  $k_L$ to $k_H$. Henceforth, the inner product definition is
\vspace{-0.2cm}
\begin{equation}
\langle \mathbf{a}, \mathbf{b}\rangle \triangleq  \sum_{k=k_L}^{k_H} a_k b_k
\vspace{-0.2cm}
\end{equation}

\vspace{-0.5cm}
\subsection{Self-Correlation}
The central idea of the proposed system is using self-correlation at the detector, rather than cross-correlation as in standard watermarking detectors. As noted in \eqref{eq:cc-reverb}, the cross-correlation with watermark template requires perfect synchronization and perfect knowledge of the acoustic channel, otherwise it will be smeared by the alternating sign of the channel response. This stringent requirement is relaxed if self-correlation is used as described in this section. 

Let $\mathbf{y}^a$ and  $\mathbf{y}^b$ be two adjacent DCT blocks of the received signal, then self-correlation is defined as 
\vspace{-0.2cm}
\begin{equation}
\psi \triangleq \langle\mathbf{y}^a, \mathbf{y}^b \rangle  \label{eq:self-corr}
\vspace{-0.2cm}
\end{equation}
The notation \emph{self-correlation} is used rather than autocorrelation to emphasize it is always between different blocks. If each block corresponds to an embedded watermarked block  as in \eqref{eq:wm_embed} after passing through the acoustic channel in \eqref{eq:chan-model}, then 
\begin{eqnarray}
\psi &=& \langle \tilde{\mathbf{x}}^a \odot {\boldsymbol \alpha} + {\n}^a , \tilde{\mathbf{x}}^b \odot {\boldsymbol \alpha} + {\mathbf{n}}^b  \rangle  \nonumber \\
   &\approx& \sum_k \alpha_k^2 x^a_k x^b_k + \sum_k \alpha_k^2 w^a_k w^b_k +  \sum_k n^a_k n^b_k  \label{eq:sc}
\vspace{-0.1cm}
\end{eqnarray}
where we assumed that the channel behavior does not change for adjacent blocks, and in the approximation we invoked the assumption of the absence of correlation between the watermark, signal, and additive noise. If the additive noise is zero-mean (which is usually the case), then the last term in \eqref{eq:sc} vanishes. If  adjacent audio blocks are weakly correlated, then the first term in \eqref{eq:sc} is much weaker than the watermark component (which is the second term in \eqref{eq:sc}), and this would improve detection. 
However, this component might become significant if a music chord is present in the host signal, and that increases the noise floor.

Note that, by employing self-correlation the impact of acoustic channel is neutralized ( by making the channel contribution nonnegative) at the cost of higher noise floor due to the host signal self-correlation. The noise-floor is significantly reduced through the sign modulation scheme that is described in the following section.

\begin{figure}
\includegraphics[scale=0.18,angle=0,trim=0mm 18mm 0mm 15mm,clip]{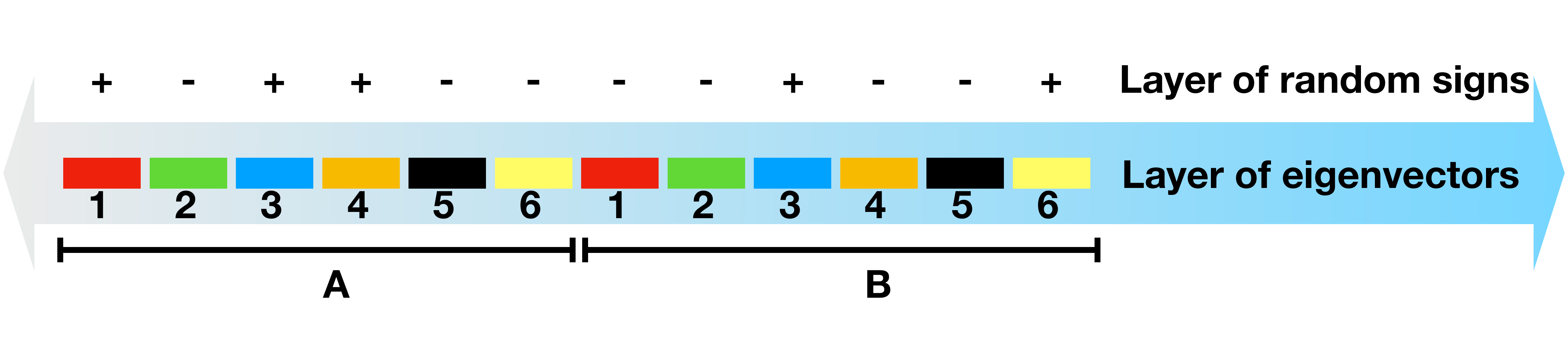}
\caption {
Illustration of the bi-layered watermark encoding structure with $N_r = 2, N_s = 6$
}\label{fig:sign-mod} 
\vspace{-0.5cm}
\end{figure}
\vspace{-0.3cm}
\subsection{Sign-Modulation Method}
The second central component in the proposed watermarking system is the sign-modulation of adjacent blocks in the host signal. 
A second encoding utilizes a sequence of $\pm 1$ to modify the binary phase of the watermark in each block. The entire encoded audio sequence can be expressed as
\vspace{-0.2cm}
\begin{equation}
\tilde \x = \bigoplus_{n=1}^{N_r} \bigoplus_{i=1}^{N_s} \Big(\x^{n,i} +  \beta\; s_{n,i}\, g_{n,i}\; \w^{i} \Big)  
\label{scencoder}
\vspace{-0.2cm}
\end{equation}
where $\bigoplus$ denotes block concatenation, $N_s$ denotes the number of segments of basic watermark building blocks, $N_r$ denotes the number of repeats of the set of segments, $\x^{n,i}$ is the $i$-th audio block of the $n$-th segment of the host audio, $\beta$ is the encoding strength, $g_{n,i} \triangleq \sqrt{\langle \x^{n,i} ,\, \x^{n,i} \rangle}$ is the segment normalization factor, and $s_{i,n}$ is an $\pm 1$ random sequence. Note that, the watermark strength at each block is proportional to the signal strength, i.e., $\eta^{n,i} = \beta g_{n,i}$, and mutually orthogonal watermarks $\{\w^i\}_{1\leq i \leq N_s}$ are inserted at each block.
An illustration of this encoding process is shown in Fig. \ref{fig:sign-mod}. Note that, each block within the segment is modulated by a random sign that will be incorporated at the decoder. Different keys could be used for the generation of the watermark and the sign sequence to allow for increased accuracy or multiple access watermarking. 

The decoder modifies the self-correlation procedure in \eqref{eq:self-corr} to accommodate multilayered embedding in \eqref{scencoder}. The multilayered self-correlation has the form
\vspace{-0.2cm}
\begin{equation}
    \label{decoder}
    \rho(t) 
        = \sum_{i=1}^{N_s} \sum_{n=1}^{N_r-1}\sum_{m=n+1}^{N_r} \frac{s_{m,i}\,s_{n,i}\,\langle \y^{m,i},\, \y^{n,i}\rangle}{h_{m,i}\; h_{n,i}}
\end{equation}
where $\rho(t)$ is the watermark decoding score, $\y^{n,i}$ is the $i$-th block of the $n$-th audio segment at the receiver, $h_{m,i} \equiv \sqrt{\langle \y^{m,i} ,\, \y^{m,i} \rangle}$,
is a normalization factor for the segment audio, $\y^{m,i}$, from the receiver.

\begin{figure}[h]
\includegraphics[scale=0.12,angle=0,trim=0mm 18mm 0mm 18mm,clip]{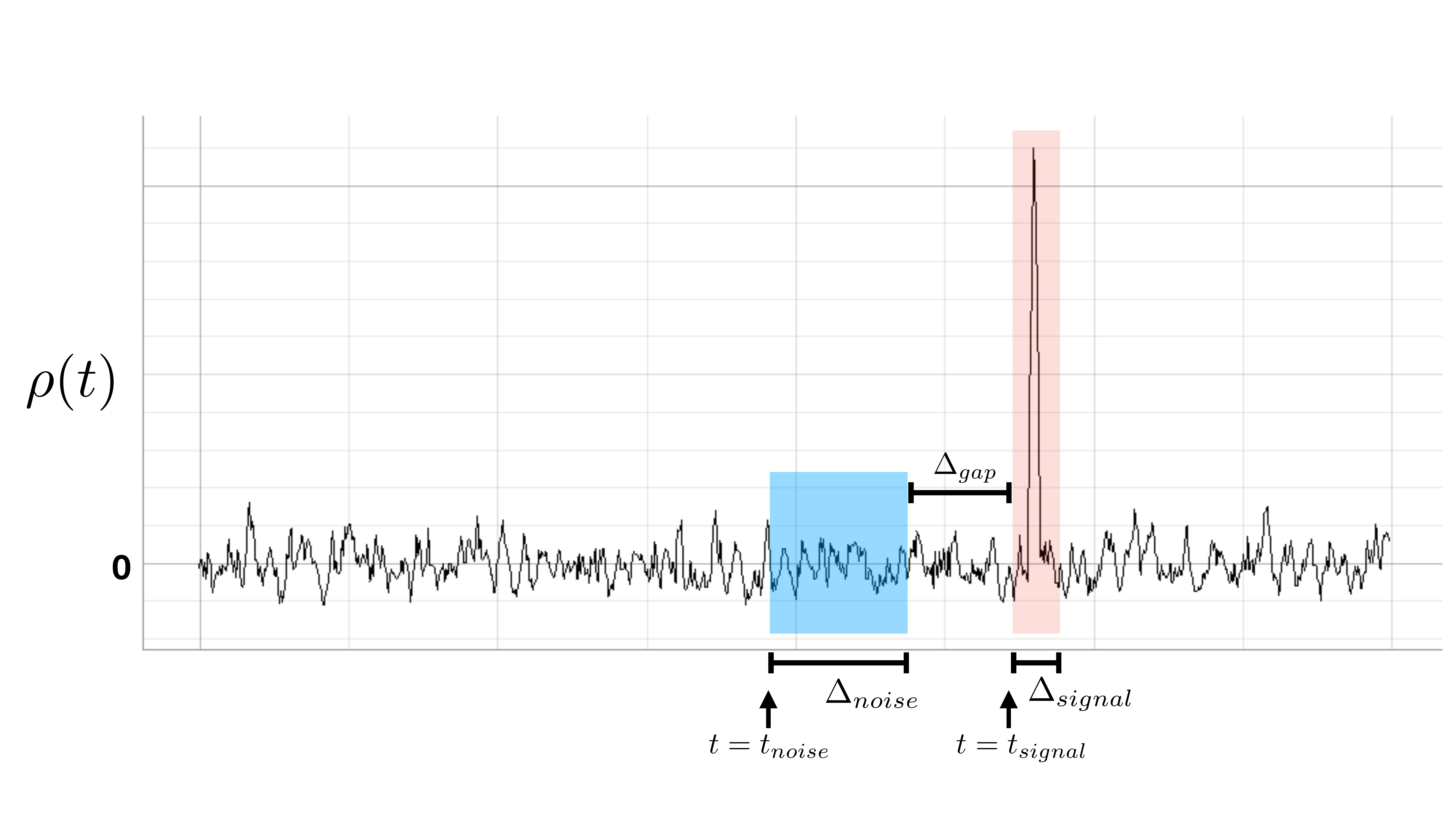}
\vspace{-0.3cm}
\caption {
Illustration of modulated self-correlation score, $\rho(t)$, under signal/null hypotheses
}\label{fig:snr-example} 
\end{figure}

Note that, with this sign modulation arrangement in the encoder and the decoder, the watermark component in \eqref{eq:sc} is invariant, while the signal component is effectively suppressed. Assuming for now that we have perfect synchronization, we will describe how $\rho$ in \eqref{decoder} behaves under signal and null hypotheses. Let
\begin{equation}
\begin{aligned}
\psi_{m,n,i} &\triangleq s_{m,i}\,s_{n,i}\,\langle \y^{m,i} ,\, \y^{n,i}\rangle \\
& = s_{m,i}\,s_{n,i}\,\langle \bx^{m,i}\odot\bdalpha+\n,\, \bx^{n,i}\odot\bdalpha+\n\rangle\\
& + \beta^2\,s_{m,i}^2\,s_{n,i}^2\, g_{m,i}\,g_{n,i}\langle \w^i\odot\bdalpha,\, \w^i\odot\bdalpha\rangle\\
& + \beta\,s_{m,i}\,s_{n,i}^2\,g_{n,i}\,\langle \bx^{m,i}\odot\bdalpha+\n,\, \w^i\odot\bdalpha\rangle\\
& + \beta\,s_{m,i}^2\,s_{n,i}\,g_{m,i}\,\langle \w^i\odot\bdalpha,\, \bx^{n,i}\odot\bdalpha+\n\rangle
\end{aligned}
\end{equation}
Note that, fractional delay of the block boundaries can be represented as part of the channel. 
Under the null hypothesis (i.e., no watermark), $\mathcal{H}_0$, i.e., when $\beta=0$,  we get the noise signature
\vspace{-0.1cm}
\begin{equation}
\rho_0(t) = \sum_{i=1}^{N_s} \sum_{n=1}^{N_r-1}\sum_{m=n+1}^{N_r}
            \frac{s_{m,i}\,s_{n,i}\,\langle \bx^{m,i}\odot\bdalpha+\n,\, \bx^{n,i}\odot\bdalpha+\n\rangle}{h_{m,i}h_{n,i}}
\end{equation}
Under the signal hypothesis (i.e., watermark exists), and after invoking the assumption of non-correlation between signal/watermark/noise, we get the signal signature (note that, $s_{m,i}^2 = s_{n,i}^2 = 1$)
\vspace{-0.1cm}
\begin{equation}
\rho_1(t) = \rho_0(t) + \sum_{i=1}^{N_s} \sum_{n=1}^{N_r-1}\sum_{m=n+1}^{N_r} \frac{\beta^2\,\langle \w^i\odot \bdalpha, \w^i\odot \bdalpha\rangle}{(h_{m,i}h_{n,i}/g_{m,i}g_{n,i})}
\end{equation}
An illustration of the modulated self-correlation behavior under both hypotheses is illustrated in Fig. \ref{fig:snr-example}. 
\vspace{-0.3cm}
\vspace{-0.1cm}

\begin{figure}[t]
\includegraphics[scale=0.7,angle=0, ,trim=0mm 6mm 0mm 1mm,clip]{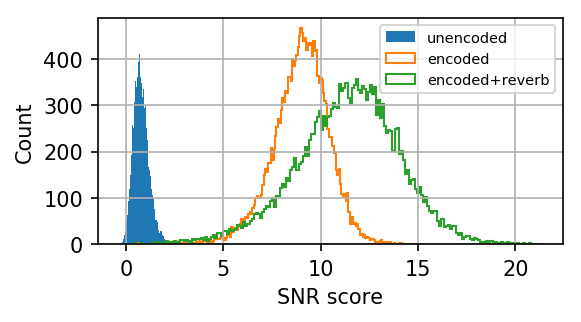}
\vspace{-0.4cm}
\center {\small{Detection Score}} 
\vspace{-0.2cm}
\caption { 
    Histogram of the detection score in \eqref{decoder} under null and signal hypotheses (with and without reverberation), averaged over 20,000 audio streams.
}\label{fig:histo-snr} 
\vspace{-0.2cm}
\end{figure}

The difficulty of decoding with self-correlation is that the mean under $\mathcal{H}_1$, and the variance under both hypotheses are dependent on the unknown channel parameters $\bdalpha$. Nevertheless, as illustrated in Fig. \ref{fig:histo-snr}, there is more than $10$ dB difference in the mean under both hypotheses, which provides flexibility to choose the detection threshold with good overall performance. In our system, the detection threshold is set to be significantly higher than the noise floor over a long period of time. Hence, the detection threshold itself is a function of the acoustic channel. 

The above discussion assumed  synchronization was achieved prior to self-correlation. In the worst case, synchronization could be achieved at a sample-level by brute-force computation of $\rho(t)$ (where fractional delay is absorbed in the channel response). Nevertheless, it was found that the modulated self-correlation mechanism tolerates imperfect alignment (roughly $\pm 50\%$ of the length of eigenvector length) with acceptable detection rate.
For example, if we take the eigenvector to be $10$ ms long, then $\pm 5$ ms misalignment can be tolerated. This is due to the blind detection procedure that parameterizes the detection parameters with the channel, and tolerable misalignments  can be modeled as part of the channel. The tradeoff between complexity and performance could be further exploited by incorporated only a subset of segments (out of $N_r$) and blocks (out of $N_s$) in \eqref{decoder}.

\vspace{-0.1cm}
\subsection{System Overview}
The overall detection procedure proceeds as follows (where $\rho(t)$ is computed as in the previous section):
\begin{enumerate}
\item Calculate the noise-mean throughout the noise region,
\begin{equation}
\vspace{-0.1cm}
    \bar\rho_{0} \equiv \frac{1}{\Delta_{n}} \sum_{t=t_{n}}^{t_{n}+\Delta_{n}} \rho(t)
\end{equation}
\vspace{-0.3cm}
\item Calculate the channel dependent noise variance,
\vspace{-0.1cm}
\begin{equation}
    \sigma_{0}^2 \equiv \frac{1}{\Delta_{n}} \sum_{t=t_{n}}^{t_{n}+\Delta_{n}-1} |\rho(t) - \bar\rho_{0}|^2
\end{equation}
\vspace{-0.3cm}
\item Set the detection threshold, $\gamma$,  at the desired point on the ROC curve, e.g.,  $\gamma = 3\sigma_0$.
\item Calculate the modulated self-correlation factor with the noise-mean correction, 
\vspace{-0.1cm}
\begin{equation}
    \bar\rho(t) \equiv \frac{1}{\Delta_{s}} \sum_{\tau=0}^{\Delta_{s}-1} \Big(\rho(t+\tau)-\bar\rho_0\Big) \label{eq:final_score}%
\end{equation}
\vspace{-0.2cm}
\item The detector operates as
\vspace{-0.1cm}
\begin{equation}
\begin{aligned}
\varepsilon(t) =
\begin{cases}
    &0, \text{\;for\;\;\;} \bar\rho(t)  <  \gamma,  \\
    &1, \text{\;for\;\;\;} \bar\rho(t) \geq  \gamma,
\end{cases}
\end{aligned}
\end{equation}
\end{enumerate}

Note that, the frequency of computing $\bar\rho(t)$ and $\varepsilon(t)$ is determined by the available computation resources.  

\section{Experimental Results}
The proposed algorithm was implemented in python for a real-time demo with consumer-grade loudspeakers and microphones. We did extensive evaluation to compute the Receiver Operating Characteristic (ROC) curve (which fully captures the detector performance \cite{kay1998fundamentals}) under different room environments and audio processing attacks.  For false accept rate calculation, we scan through a non-watermarked audio of duration $\sim 41$ min  every 5 milliseconds.
For the detection part, the watermark is inserted every $4$ seconds in the same host audio, i.e., $\sim 600$ watermarks. 

The system was first evaluated versus standard audio processing operations, e.g., lowpass filtering, highpass filtering, and mp3 compression. It showed the standard robust performance of spread spectrum systems \cite{kirovski2003spread}. The subjective quality of the watermarked audio was evaluated by $10$ expert listeners and was shown to be indistinguishable from original audio. Both the encoder and the decoder run in real-time, and the latency is only due to audio block buffering delay.

\begin{figure}[t]
\includegraphics[scale=0.18,angle=0]{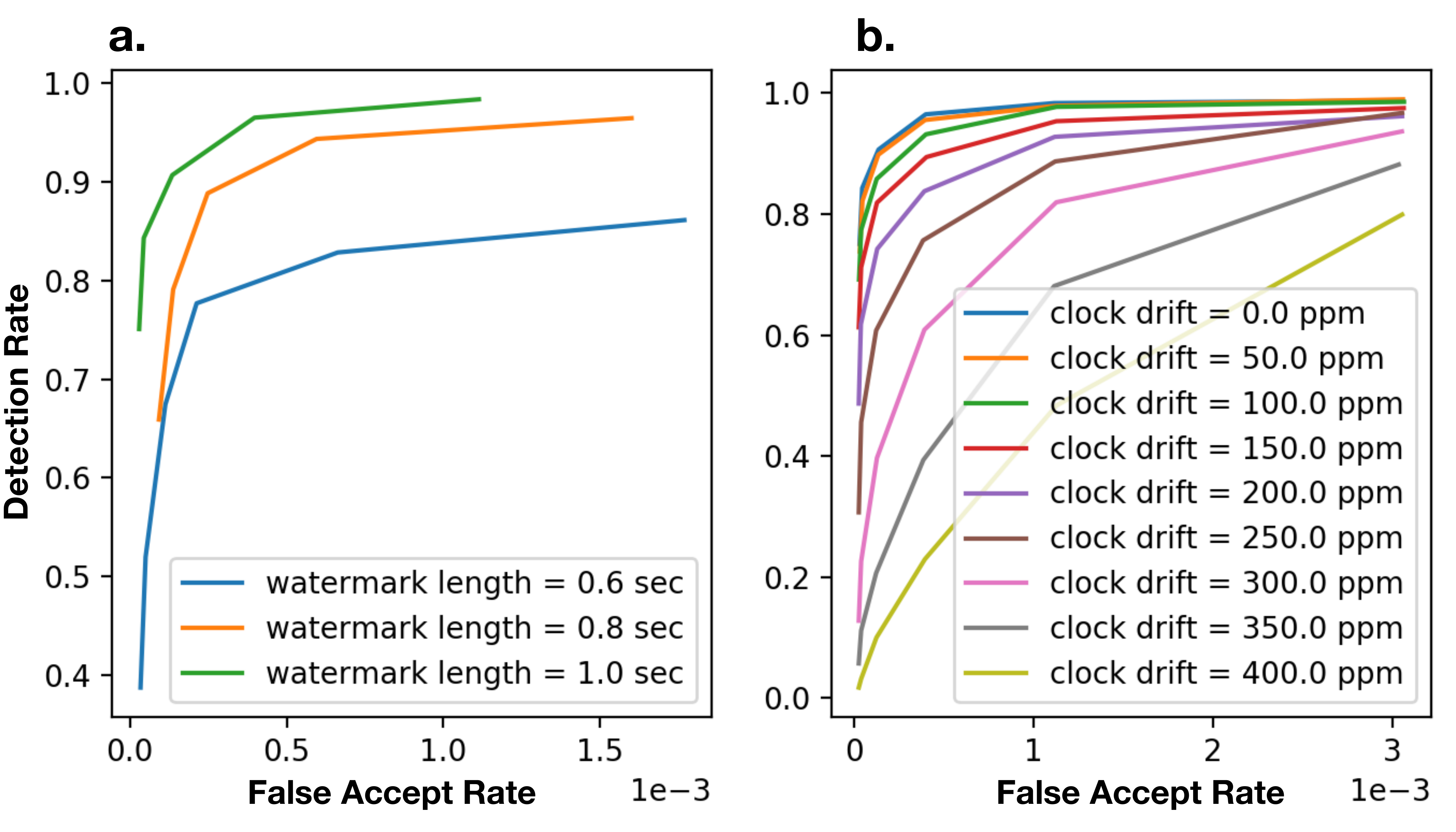}
\caption { 
(a.) The ROC with reverberation versus different watermark length where we set $N_s=2$;
(b.) The ROC with reverberation and clock drift (in ppm), the watermark length is fixed to 1 sec with $N_s=2$.
}\label{fig:ROC_wm} 
\vspace{-0.4cm}
\end{figure}

Next, we evaluate the robustness of the proposed system to room reverberation. The ROC curve is computed under $12$ different room conditions, and the results are averaged for different sizes of the embedded watermark, which is also proportional to the overall system latency. 
Multiple watermarks with different duration are simultaneously inserted in the host audio. This has a minor impact because the watermarks are mutually orthogonal. In evaluating the ROC curve, we applied measured reverberation filters to the watermarked audio prior to the detector. 
Fig. \ref{fig:ROC_wm}a shows the ROC for different watermark performance versus watermark length (with no clock drift between the encoder and the decoder). In the figure, we zoomed in the horizontal axis of the ROC curve, because of the almost perfect behavior when the watermark duration is longer than $0.8$ second.

Finally, we evaluated the combined impact of clock drift and reverberation. In this experiment, both the encoder and the decoder run at the same sampling frequency, but the decoder clock is perturbed by different ppm values, and the decoder is run without clock correction. The resulting ROC behavior is shown in Fig. \ref{fig:ROC_wm}b, where we also zoomed in the horizontal axis to clarify the behavior. As noted from the figure, the performance is robust to clock drift up to  $\sim 300$ ppm, when the watermark duration is $1$ second.

\section{Conclusion}
With the complicated indoor acoustic channel, there are two choices for successful audio watermarking. The first choice is to estimate and equalize the channel prior to applying a standard detector. The second choice is to restructure the detector to neutralize the channel impact without explicit channel estimation. The first choice is optimal from performance perspective, but it is expensive  in computation and latency. The second choice, which we adopted in this work, is more appropriate for real-time embedded system despite its suboptimal detector. The suboptimal performance could always be enhanced by using a longer or stronger watermark. The proposed novel detector utilizes modulated self-correlation between adjacent audio blocks, which effectively neutralizes the indoor channel impact and eliminates the need for explicit channel estimation. The experimental results showed the robustness of the system when the watermark duration is greater than $0.8$ second under general reverberation conditions, and with clock drift up to $300$ ppm.


In a future work, we describe a dynamic programming algorithm to prune the synchronization search by an order of magnitude with negligible impact on the detection performance. Future work also includes deploying microphone arrays \cite{chhetriAFE2018Eusipco} to improve self-correlation, and reduce the reverberation impact.

\bibliographystyle{IEEEbib}
\bibliography{refs}

\begin{thebibliography}{10}

\bibitem{swanson1998audio}
Mitchell~D Swanson, Bin Zhu, and Ahmed~H Tewfik,
\newblock ``Audio watermarking and data embedding--current state of the art,
  challenges and future directions,''
\newblock in {\em Multimedia and Security Workshop at ACM Multimedia}.
  Citeseer, 1998, vol.~41.

\bibitem{hua2016twenty}
Guang Hua, Jiwu Huang, Yun~Q Shi, Jonathan Goh, and Vrizlynn~LL Thing,
\newblock ``Twenty years of digital audio watermarking?a comprehensive
  review,''
\newblock {\em Signal Processing}, vol. 128, pp. 222--242, 2016.

\bibitem{cesar2009leveraging}
Pablo Cesar, Dick~CA Bulterman, and Jack Jansen,
\newblock ``Leveraging user impact: an architecture for secondary screens usage
  in interactive television,''
\newblock {\em Multimedia systems}, vol. 15, no. 3, pp. 127--142, 2009.

\bibitem{kuttruff2016room}
Heinrich Kuttruff,
\newblock {\em Room acoustics},
\newblock Crc Press, 2016.

\bibitem{mansour2003time}
Mohamed~F Mansour and Ahmed~H Tewfik,
\newblock ``Time-scale invariant audio data embedding,''
\newblock {\em EURASIP Journal on Applied Signal Processing}, vol. 2003, pp.
  993--1000, 2003.

\bibitem{pun2013robust}
Chi-Man Pun and Xiao-Chen Yuan,
\newblock ``Robust segments detector for de-synchronization resilient audio
  watermarking,''
\newblock {\em IEEE Transactions on Audio, Speech, and Language Processing},
  vol. 21, no. 11, pp. 2412--2424, 2013.

\bibitem{wang2006novel}
Xiang-Yang Wang and Hong Zhao,
\newblock ``A novel synchronization invariant audio watermarking scheme based
  on dwt and dct,''
\newblock {\em IEEE Transactions on signal processing}, vol. 54, no. 12, pp.
  4835--4840, 2006.

\bibitem{xiang2014patchwork}
Yong Xiang, Iynkaran Natgunanathan, Song Guo, Wanlei Zhou, and Saeid Nahavandi,
\newblock ``Patchwork-based audio watermarking method robust to
  de-synchronization attacks,''
\newblock {\em IEEE/ACM Transactions on Audio, Speech, and Language
  Processing}, vol. 22, no. 9, pp. 1413--1423, 2014.

\bibitem{nadeau2017audio}
Andrew Nadeau and Gaurav Sharma,
\newblock ``An audio watermark designed for efficient and robust
  resynchronization after analog playback,''
\newblock {\em IEEE Transactions on Information Forensics and Security}, vol.
  12, no. 6, pp. 1393--1405, 2017.

\bibitem{del2011audio}
Giovanni Del~Galdo, Juliane Borsum, Tobias Bliem, Alexandra Craciun, and Stefan
  Kr{\"a}geloh,
\newblock ``Audio watermarking for acoustic propagation in reverberant
  environments,''
\newblock in {\em Acoustics, Speech and Signal Processing (ICASSP), 2011 IEEE
  International Conference on}. IEEE, 2011, pp. 2364--2367.

\bibitem{doubledct}
Xia Zhang, Di~Chang, Wanyi Yang, Qian Huang, Wei Guo, and Yanbin Zhao,
\newblock ``An audio digital watermarking algorithm transmitted via air channel
  in double dct domain,''
\newblock in {\em Multimedia Technology (ICMT), 2011 International Conference
  on}. IEEE, 2011, pp. 2926--2930.

\bibitem{cox1996secure}
Ingemar~J Cox, Joe Kilian, Tom Leighton, and Talal Shamoon,
\newblock ``Secure spread spectrum watermarking for images, audio and video,''
\newblock in {\em Image Processing, 1996. Proceedings., International
  Conference on}. IEEE, 1996, vol.~3, pp. 243--246.

\bibitem{kirovski2003spread}
Darko Kirovski and Henrique~S Malvar,
\newblock ``Spread-spectrum watermarking of audio signals,''
\newblock {\em IEEE transactions on signal processing}, vol. 51, no. 4, pp.
  1020--1033, 2003.

\bibitem{PLLbook}
Roland~E Best,
\newblock {\em Phase locked loops: design, simulation, and applications},
\newblock McGraw-Hill Professional, 2007.

\bibitem{martucci1994symmetric}
Stephen~A Martucci,
\newblock ``Symmetric convolution and the discrete sine and cosine
  transforms,''
\newblock {\em IEEE Transactions on Signal Processing}, vol. 42, no. 5, pp.
  1038--1051, 1994.

\bibitem{matrixbook}
Roger~A Horn, Roger~A Horn, and Charles~R Johnson,
\newblock {\em Matrix analysis},
\newblock Cambridge university press, 1990.

\bibitem{kay1998fundamentals}
Steven~M Kay,
\newblock ``Fundamentals of statistical signal processing, vol. ii: Detection
  theory,''
\newblock {\em Signal Processing. Upper Saddle River, NJ: Prentice Hall}, 1998.

\bibitem{chhetriAFE2018Eusipco}
Amit Chhetri, Philip Hilmes, Trausti Kristjansson, Wai Chu, Mohamed Mansour,
  Xiaoxue Li, and Xianxian Zhang,
\newblock ``Multichannel {A}udio {F}ront-{E}nd for {F}ar-{F}ield {A}utomatic
  {S}peech {R}ecognition,''
\newblock in {\em 2018 European Signal Processing Conference (EUSIPCO)}, 2018,
  pp. 1541--1545.

\end{thebibliography}

\end{document}